\documentclass[%
  twocolumn,%
  groupedaddress,%
  showpacs,%
  prb,%
  letterpaper,%
  amsfonts,%
]{revtex4}

\usepackage[nolist,nohyperlinks]{acronym}

\usepackage{amsmath,amssymb,bm}
\usepackage{graphicx,subfigure}
\usepackage{hyperref}
\hypersetup{%
  breaklinks = {true},
  citecolor = {blue},
  colorlinks = {true},
  linkcolor = {red},
  pdfauthor = {\textcopyright\ Vitor M. Pereira},
  pdfcreator = {\LaTeX\ and \flqq hyperref\frqq},
  pdffitwindow = {true},
  pdfmenubar = {true},
  pdfpagelayout = {SinglePage},
  pdfstartview = {Fit},
  pdftoolbar = {true},
  plainpages = {false},
}

\newcommand{\vF}{\ensuremath{v_\text{F\,}}}
\newcommand{\bsigma}{\ensuremath{\bm\sigma}}

\newcommand{\br}{\ensuremath{\bm {r}}}
\newcommand{\Hcal}{\mathcal{H}}

\newcommand{\Ncal}{\mathcal{N}}

\newcommand{\gc}{g_\text{c}}
\newcommand{\gct}{\tilde{g}_\text{c}}
\newcommand{\e}{\epsilon}

\newcommand{\kt}{\tilde{k}}

\newcommand{\Eq}{Eq.}
\DeclareMathOperator{\sign}{sign}

\begin{document}


\title{Supercritical Coulomb Impurities in Gapped Graphene}

\pacs{81.05.Uw,71.55.-i,25.75.Dw}

%

\author{Vitor M. Pereira}
\affiliation{Department of Physics, Boston University, 590
Commonwealth Avenue, Boston, MA 02215, USA}

\author{Valeri N. Kotov}
\affiliation{Department of Physics, Boston University, 590
Commonwealth Avenue, Boston, MA 02215, USA}

\author{A.~H. Castro Neto}
\affiliation{Department of Physics, Boston University, 590
Commonwealth Avenue, Boston, MA 02215, USA}

\date{\today}


\begin{abstract}
We study the problem of Coulomb field-induced charging of the ground state in a
system of 2D massive Dirac particles -- gapped graphene.
As in its 3D QED counterpart, the critical Coulomb coupling is renormalized to
higher values, compared to the massless case. 
We  find that  in gapped graphene a novel supercritical regime is possible,
where the screening charge is comparable to the impurity charge,
thus leading to suppression of the Coulomb field at nanometer scales. 
We corroborate this with numerical solution of the tight-binding
problem on the honeycomb lattice.
\end{abstract}

\maketitle


%
%
\section{Introduction}

The successful experimental isolation of a single plane of
$sp^2$ carbon atoms (graphene) has stirred our deepest assumptions about
condensed matter systems \cite{Novoselov:2005,CastroNeto:2006b,Geim:2007}. 
Not only is graphene unique in the realm of solid state, but it has also a vast
potential as a test ground for many of the remarkable
predictions of \ac{QED} \cite{Neto:2006,Katsnelson:2007}. 
One remarkable characteristic of graphene is that its electronic properties can
be tailored in many different ways, either by applying transverse electric and
magnetic fields, changing its geometry, or modifying the substrate where
graphene is deposited or grown \cite{Geim:2007}.

In fact, while graphene deposited in SiO$_2$ is well described by the \ac{2D}
massless Dirac equation \cite{Novoselov:2005}, graphene
grown on SiC can be described in terms of massive \ac{2D} Dirac electrons
\cite{Zhou:2007}.
Substrate induced potentials can break symmetries of the
honeycomb lattice and generate gaps in the electronic spectrum. Hence,
by suitable choice of substrates one can tune the ``rest mass'' of the
``relativistic'' particles and explore new phenomena beyond the massless case. 
Furthermore, the electronic properties of devices, such as
carrier mobility, depend strongly on how the electronic degrees of freedom 
(massive or massless) interact with impurities. Of particular importance are
charged impurities that naturally appear either on the substrate, on top of
graphene, or between the substrate and graphene. 
Charged impurities play an important role in the transport characteristics of
graphene deposited in SiO$_2$ \cite{Nomura:2007,Adam:2007,Chen:2008} and should
play an important role in epitaxial graphene as well \cite{Heer:2007}. 

In this paper we explore the non-trivial restructuring and charging of the
vacuum that occurs in the presence of a supercritical Coulomb center
\cite{Zp:1972,Greiner:1985} if the system is described by a ``massive''
spectrum. 
The problem of an unscreened Coulomb impurity in undoped graphene has
recently received considerable attention
\cite{Shytov:2007ab,Novikov:2007,Pereira:2007,Biswas:2007,Fogler:2007,
Terekhov:2007}. 
This is justified since, on the one hand, the vanishing \ac{DOS} at the Fermi
energy of undoped graphene and the absence of back-scattering suppress screening
\cite{Ando:2006}. In addition, the Coulomb problem in
graphene is  the condensed matter analogue of supercritical
nuclei ($Z\alpha>1$) in \ac{QED}, whose rich and fundamental phenomena remain
elusive to experimental testing \cite{Shytov:2007ab,Pereira:2007,Greiner:1985}.
We show here that this analogy achieves its fullest in gapped graphene, where
one can resolve the incremental charging of the vacuum, with strong
implications for the screening of the Coulomb center.

The rest of the paper is organized as follows.
In Sec.~\ref{sec:Model} we introduce  our model, and in Sec.~\ref{sec:Massless}
its properties in the massless limit are reviewed.  
In Sec.~\ref{sec:Massive} the behavior of discrete energy levels in the massive
case is discussed. 
We examine in detail the structure of the critical wavefunction and the
corresponding renormalization of the critical Coulomb coupling in
Sec.~\ref{sec:CriticalWavefunction}. In Sec.~\ref{sec:VacuumPolarization} we
study the behavior of the vacuum charge across the critical point.
Sec.~\ref{sec:FiniteSystem} contains the corresponding results for a finite-size
system (where the energy gap is due to the finite size only).
Sec.~\ref{sec:Conclusions} contains a discussion of our results and conclusions.

%
%
\section{The Model}\label{sec:Model}
To address the Coulomb problem in graphene we resort to the single-valley
effective Dirac description of the electron dynamics. We start from the
tight-binding Hamiltonian describing electrons in a honeycomb lattice, and in
the presence of a Coulomb center of strength $Z$. In addition, we assign a
different local energy to each sublattice. The resulting Hamiltonian is 
\begin{align}
   \Hcal &= t \sum_{i}
    \bigl(a^\dagger_i b_i + \text{h.c.}\bigr) -
      Z e^2 \sum_{i}
      \biggl(
        \dfrac{a^\dagger_i a_i}{r^A_i} + \dfrac{b^\dagger_i b_i}{r^B_i}
      \biggr) 
    \nonumber\\
    &+  \frac{\Delta}{2} \sum_i a^\dagger_i a_i -
        \frac{\Delta}{2}\sum_i b^\dagger_i b_i
  \,.
  \label{eq:H-Lattice}
\end{align}
In the above, the sums run over unit cells, and the operators $a_i$($b_i$)
pertain to the A(B) sublattice.
This system exhibits a band gap of $\Delta$ and, if undoped, has an insulating
ground state with the Fermi level in the gap.
Just as in its gapless counterpart, low energy excitations can be addressed
within an effective mass approximation, consisting of a $\bm{k}.\bm{p}$
expansion around the $K$ and $K'$ points in the Brillouin zone. The
resulting effective Hamiltonian leads to the Dirac equation in 2D under a
Coulomb field:
\begin{equation}
  \Bigl(
    -i \hbar\vF \bsigma\cdot \bm{\nabla} - \dfrac{Z e^2}{r} +
    m\vF^2 \sigma_z
  \Bigr)
  \Psi(\br) = \e\;\Psi(\br)
  \label{eq:SchrodingerEq}
  \,,
\end{equation}
where the wavefunction $\Psi({\br})$ has a spinor structure that
carries the amplitude of the wavefunction on each sublattice,
\begin{equation}
  \Psi(\br) = \binom{\psi^A(\br)}{\psi^B(\br)}
  \,.
\end{equation}
$\vF = 3 t a/ (2 \hbar)$ is the Fermi velocity in graphene and $a$ is the C--C
distance.
It is convenient to introduce $g$ as the dimensionless coupling to the external
Coulomb field: $g \equiv Z \alpha$, and $\alpha\equiv e^2/(\hbar \vF)$ is the
``fine structure'' constant of graphene. 
The gap in the spectrum is induced by the term proportional to $\sigma_z$, akin
to a relativistic rest mass. The mass $m$ is related to the local energy
mismatch in the tight-binding formulation through $\Delta = 2m\vF^2$. To avoid
too cumbersome a notation, throughout the paper we shall take a system of units
in which $\hbar=\vF=1$. Without loss of generality, the impurity will be
considered attractive ($g>0$) and the C-C distance ($a\simeq 1.42 \text{\AA}$)
will be used as the length unit.

In parallel with the exact analytical solution of Eq.~\eqref{eq:SchrodingerEq},
we solve the tight-binding problem of \eqref{eq:H-Lattice} exactly, via
direct numerical diagonalization on the lattice. This exact solution provides
an important control of the validity of the Dirac approach, and the results of
the two approaches will be frequently compared.

%
%
\section{Coulomb Impurities in Massless Graphene}%
\label{sec:Massless}
Before delving into the details of the supercritical regime in the massive case,
a brief overview of the main physics seen in the massless problem is pertinent
\cite{Shytov:2007ab,Pereira:2007,Novikov:2007}.
When $m=0$ Eq.~\eqref{eq:SchrodingerEq} separates in cylindrical
coordinates \cite{DiVincenzo:1984}, and can be
subsequently mapped into the radial Schr\"odinger equation of the usual Coulomb
problem in 3D \cite{Pereira:2007}. 
One decisive peculiarity of this mapping is that, although the radial equation
is formally the same as the radial equation in the Schr\"odinger Coulomb
problem, the angular momentum quantum number appears replaced by an
``effective angular momentum'' equal to $l=\sqrt{j^2-g^2}$. Here, $j$ is
the quantum number associated with the total angular momentum operator
\begin{equation}
  J_z=L_z + \frac{1}{2}\sigma_z
  \,,
\end{equation}
and takes the values $j=\pm1/2,\pm3/2$,\ldots.
The radial solutions are thus expressed in terms of the Coulomb functions
\cite{Abramowitz:1964,Pereira:2007} $F_l(-g \sign(\e), |\e| r)$ and 
$G_l(-g \sign(\e), |\e| r)$. Given that, as usual, $l$ determines the asymptotic
power law behavior of the wavefunctions:
\begin{equation}
  F_l(r\simeq 0) \sim r^{l+1}\,,\qquad
  G_l(r\simeq 0) \sim r^{-l}\,,\nonumber
\end{equation}
it is evident that $g=\gc=1/2$ is a singular point for the lowest total angular
momentum channel ($j=\pm 1/2$). Below $\gc$ the space of solutions is
constrained by the requirement of regularity at the origin, as usual, and this
selects $F_l(r)$ as the regular solution. But above $\gc$, $l$ becomes imaginary
and any linear combination of the
solutions ($F_l, G_l$) becomes square integrable at the origin. At the same
time, one sees from the above asymptotics that the solutions oscillate endlessly
when $r\to 0$ as $\propto \exp[i\log(r)]$.
This is a signature of the ``fall
to the center'' characteristic of highly singular potentials
\cite{Landau-QM:1981}. In fact, the supercritical regime can be
intuitively understood from a classical perspective: consideration of the
classical equations of motion for the Dirac Hamiltonian shows that, for a given
angular momentum, there will be a critical coupling above which the classical
orbits spiral and fall onto the potential source \cite{Shytov:2007ab}. The
potential has become too singular and there are no closed nor scattering orbits.

It is known that, in this case, the quantum mechanical problem becomes uniquely
defined only after an additional boundary condition is introduced, reflecting
the physical cutoff of the potential at short distances
\cite{Case:1960,Zp:1972}.
For graphene the natural cutoff is the lattice spacing, $a$. 
This regularization of the potential at short distances permits the exact
solution to be extended to the supercritical regime ($g>\gc$) which, among 
other effects, is characterized by the
presence of marked resonances in the spectral density of the hole
channel \cite{Pereira:2007}. Such spectral features are analogous to the 
positron resonances expected for a supercritical nucleus in
\ac{QED} \cite{Greiner:1985}. But a fundamental difference exists between the
two cases: whereas in \ac{QED}, due to the finite mass,  the emergence of
positron resonances is an incremental process (as a function of $g$), in
massless graphene an infinite number of them instantly appears at $g=\gc$. 
Semiclassical considerations illustrate how these
resonances emerge from an infinite number of quasi-bound states
embedded in the lower continuum\cite{Shytov:2007ab}.
Adding to these spectral peculiarities, the induced electronic charge behaves
rather differently on the two sides of the critical point, being localized close
to the impurity for $g<\gc$, and otherwise decaying algebraically
\cite{Shytov:2007ab,Pereira:2007} as $1/r^2$.

An important question is how will these non-perturbative effects contribute to
screen the impurity potential. 
Adding to the ever present virtual vacuum polarization arising from
virtual excitations \cite{Terekhov:2007}, in the supercritical regime
one expects those quasi-localized states to partially screen the Coulomb
center. It was argued, on the basis
of a self-consistent treatment, that the impurity charge at large distances is
screened down to its critical value $Z_c=1/(2\alpha)$ for
any $g>\gc$ \cite{Shytov:2007ab}. 
We will show that in the massive case of interest here the situation
is conceptually different, and a novel regime can be reached where the
polarization charge is comparable to the bare charge $Z$. This leads to a
strong tendency to 
neutralization of the impurity potential at a length scale set by the Compton
wavelength in graphene. We examine in detail the conditions for this to occur.

%
%
\section{Massive Dirac Fermions in Graphene}%
\begin{figure}[tb]
  \centering
  \subfigure[][]{%
    \includegraphics*[clip, width=0.58\columnwidth]{%
      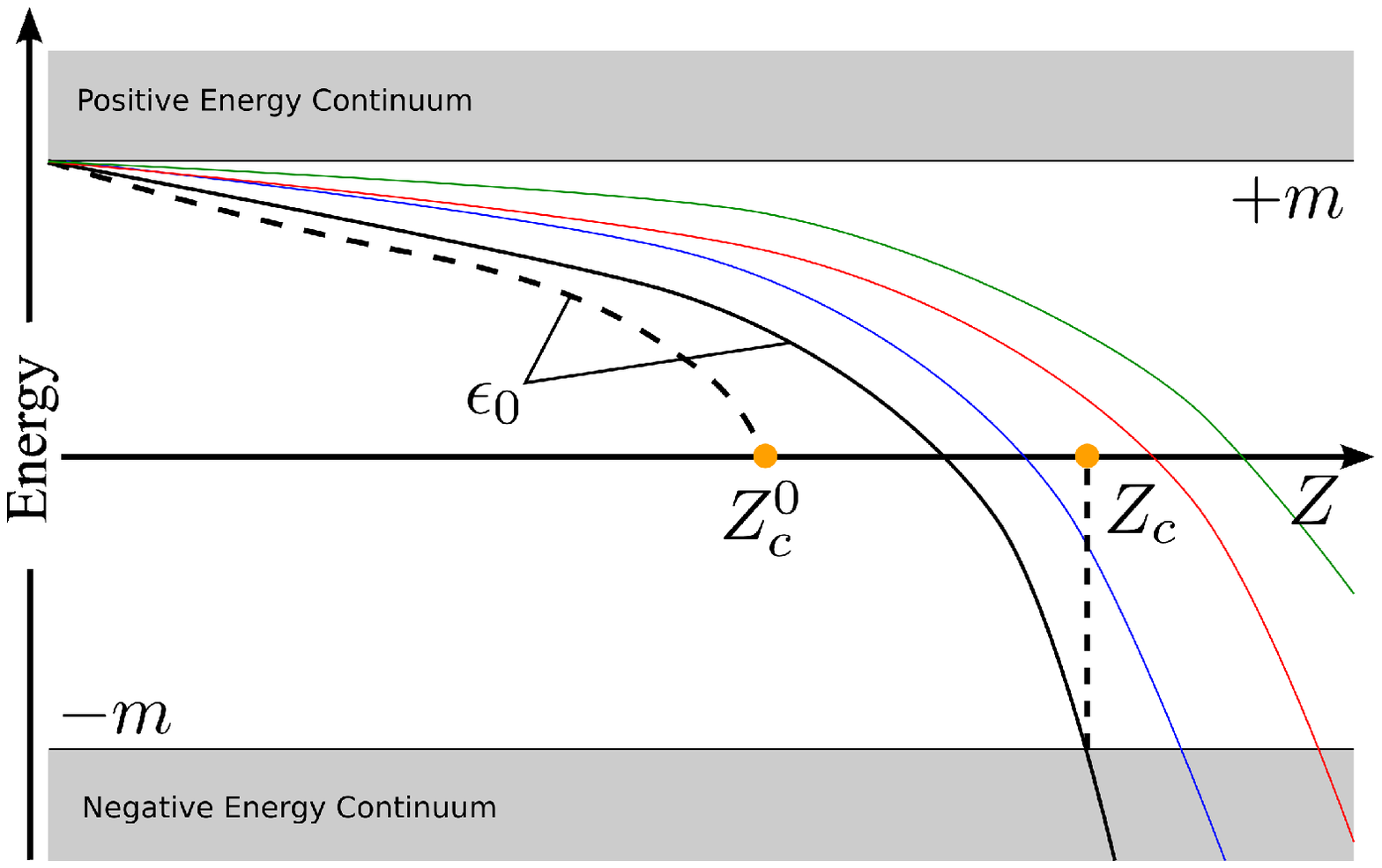%
    }
    \label{fig:Diving}%
  }%
  \subfigure[][]{%
    \includegraphics*[clip, width=0.35\columnwidth]{%
      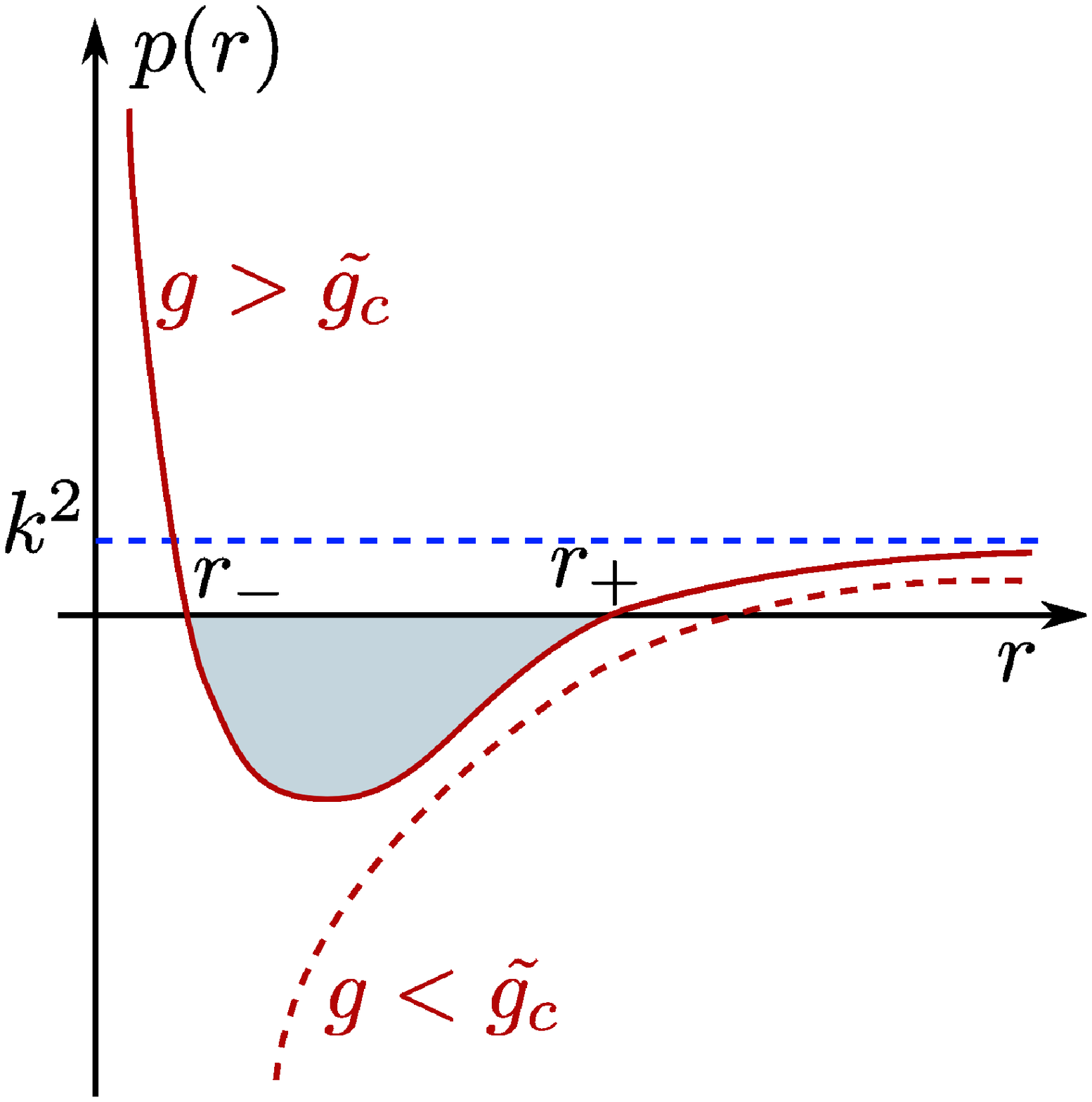%
    }
    \label{fig:Semiclassics}%
  }%
  \caption{
    \subref{fig:Diving} Schematic depiction of the change of bound levels
    \eqref{eq:FineStructure} with increasing coupling. 
    The dashed line represents the evolution of $\e_0$ for a point nucleus,
    and the solid one is for a regularized potential. Adapted from
    Ref.~\onlinecite{Muller:1975}.
    \subref{fig:Semiclassics} Effective semiclassical momentum of
    Eq.~\eqref{eq:SemiclassicalMomentum} when 
    $\e_0 \approx -m$ and $g>\gct$ (solid) or $g<\gct$ (dashed).
  }
  \label{fig:Schematics}
\end{figure}

\label{sec:Massive}
The emergence of resonant solutions described above is best appreciated when
the system acquires a mass and the electron dispersion becomes
$\e^2=k^2+m^2$.
This mass can arise  physically  from  sublattice symmetry breaking
\cite{Zhou:2007}, spin-orbit coupling or from reduced dimensionality, as in a
finite-size mesoscopic sample. 
The one-particle solution for the massive case is straightforward and is
discussed, for instance, in Refs.~\onlinecite{Novikov:2007,Khalilov:1998}. The
fundamental difference that a finite mass brings to the Coulomb problem is the
immediate appearance of bound solutions corresponding to the hydrogen-like fine
structure spectrum in 2D:
\begin{equation}
  \e_{n,j} = m \sign(g)
    \frac{n+\sqrt{j^2-g^2}}{\sqrt{g^2 + \bigl[n+\sqrt{j^2-g^2}\bigr]^2}}
  \label{eq:FineStructure}
  \,.
\end{equation}
In the above $n$ is the principal quantum number and $j$ the total angular
momentum quantum number defined above. These states reside in the gap $-m <
\epsilon < m $, and are described by wavefunctions that decay exponentially for
distances larger than the ``Bohr'' radius, $a_0=\lambda_C/(Z \alpha)$, where
$\lambda_C = \hbar/(m \vF)$ is the ``Compton'' wavelength in graphene. 
For an attractive potential the lowest bound state ($n=0,j=1/2$) has an energy
given by 
\begin{equation}
  \epsilon_0 = m \sqrt{1 - (2 g)^2}
  \,,
  \label{eq:GSenergy}
\end{equation}
and hence reaches zero  singularly  at $g=\gc$, becoming imaginary beyond this
coupling strength. Similarly to what we have discussed regarding the massless
case, this imaginary energy reflects the
fact that a description of a point nucleus is impossible for $g>\gc$. A
physical regularization procedure, where one cuts the potential off at small
$r$, relaxes this constraint and the bound levels are then allowed to sink
further \cite{Popov:1971,Greiner:1985} through negative energies until the point
$\epsilon=-m$ is reached. The diving of the bound solutions with
increasing coupling is schematically depicted in Fig.~\ref{fig:Diving}.
The vicinity of this point, where the lowest bound state is about to merge (or
has just merged) with the continuum is the focus of all our subsequent
discussion.

%
%

%
%
\section{Critical Wavefunction and Critical Coupling Renormalization}%
\label{sec:CriticalWavefunction}
Given that in a regularized Coulomb potential the bound levels are allowed to
dive into negative energies, the relevance is transferred from the
particular coupling $g=\gc=1/2$ to the value $\gct$ at which $\epsilon_0=-m$.
Necessarily $\gct>\gc$, and the critical coupling is thus renormalized to higher
values. This is one of the immediate consequences of the presence of a gap. 
The merging of the bound solution with the continuum is nonetheless
peculiar, and hence is worth carefull consideration. We begin with a
semiclassical argument. The first important thing to notice is that the merging
state does not become completely delocalized, as one would expect in
typical Schr\"odinger problems, but remains localized as it dives into the lower
continuum. This is somewhat counter-intuitive and is best appreciated by
considering the problem semiclassically within WKB. An alternative to the
full WKB approximation is to consider the \emph{``classical relativistic''}
momentum
\begin{equation}
  p(r)^2 = \bigl[\e - U(r)\bigr]^2 - m^2
  \,
\end{equation}
where $U(r)$ is the Coulomb potential. With this definition, the radial momentum
$p_r$ is written as 
\begin{equation}
  p_r(r)^2 = k^2 + 2g\e/r + (g^2-\ell^2)/r^2
  \label{eq:SemiclassicalMomentum}
  \,,  
\end{equation}
with $\ell$ the classical effective angular momentum and $k^2=\e^2-m^2$.
The solid line in Fig.~\ref{fig:Semiclassics} shows the profile of
$p_r(r)^2$ slightly after the merging has taken place. There is a classically
forbidden region [$p(r)^2<0$] between $r_-$ and $r_+$, and the long range
Coulomb tail implies that $r_+\gg r_-$. In fact, precisely at $\e_0 = -m$ we
have $r_+ = \infty$, and the barrier is infinite. 
Close to the critical point we can write $\e_0 \simeq -m -\beta (g-\gct)$ and
examine the penetrability of this barrier within WKB. It reads
\begin{equation}
  w = e^{-2S_\text{cl}} = 
  \exp\Biggl(-\frac{2\pi m \gct}{\sqrt{2 \beta m}\sqrt{g-\gct}}\Biggr)
  \label{eq:Penetrability}
  \,,
\end{equation}
and becomes exponentially small as $g\to\gct^+$. Such wide
barrier (absent when $\e_0\simeq +m$) ensures that the state remains
appreciably localized, even after merging into the lower continuum.

We now turn to the exact quantum mechanical solution of
Eq.~\eqref{eq:SchrodingerEq} for the particular regularization of the Coulomb
potential described by \cite{Popov:1971}
\begin{equation}
  V(r) = \begin{cases}
           - g/r, & r > a\\
           - g/a, & r \le a
         \end{cases}
  \,.
  \label{eq:Regularization}
\end{equation}
This regularization represents the physical situation in which there is a
Coulomb impurity at the center of an hexagon in the honeycomb lattice. In this
case the lattice parameter, $a$, is the closest distance that an electron
hopping among carbon sites can be from the potential source. It is also a good 
approximation for an impurity placed slightly above the graphene plane.
Cylindrical symmetry is preserved by this regularization and
Eq.~\eqref{eq:SchrodingerEq} naturally separates in cylindrical coordinates.
We define the radial spinor components as ($j=\pm1/2,\pm3/2,\dots$)
\begin{equation}
  \Psi_j(r,\varphi) = 
    \frac{1}{\sqrt{r}}
    \begin{pmatrix}
      e^{-i (j - 1/2)\varphi} A(r) \\
      i e^{-i (j + 1/2)\varphi} B(r) 
    \end{pmatrix}
  \,,
  \label{eq:SpinorDef}
\end{equation}
after which the radial equations for \eqref{eq:SchrodingerEq} become $(r>a)$
\begin{equation}
  \begin{bmatrix}
    (\e + \frac{g}{r} - m) & -(\partial_r + \frac{j}{r}) \\   
    (\partial_r - \frac{j}{r}) & (\e + \frac{g}{r} + m) 
  \end{bmatrix}
  \begin{bmatrix}
    A(r)\\
    B(r)    
  \end{bmatrix}
  = 0
  \,.
  \label{eq:RadialEq}  
\end{equation}
We will be  interested in the states at the threshold, with energy $\e=-m$. In
that case, the equation for the $A(r)$ component reads ($r>a$)
\begin{equation}
  r^2 \bigl( \sqrt{r} A(r) \bigr)'' + 
  \Bigl[ g^2 - j^2 + \frac{1}{4} - 2 g m r \Bigr] \sqrt{r} A(r) =0
  \,,
\end{equation}
whose solution is given in terms of the modified Bessel functions  $I_\nu(z)$
and $K_\nu(z)$. By imposing a vanishing boundary condition at infinity one 
obtains ($\Ncal$ is a normalization constant),
\begin{subequations} \label{eq:ABsolOutside}
\begin{equation}
  A(r) = \Ncal K_{i\nu}(\sqrt{8gmr})
  \label{eq:AsolOutside}
  \,,
\end{equation}
where $\nu=2\sqrt{g^2-j^2}$. 
We note that this solution has the same form as the corresponding spinor
component in the 3D \ac{QED} problem \cite{Popov:1971,Zp:1972}. The B component
follows  directly from \eqref{eq:RadialEq}:
\begin{align}
  B(r) = \frac{\Ncal}{g}
    \biggl[ &
      \Bigl(j+\frac{i\nu}{2}\Bigr) K_{i\nu}\bigl(\sqrt{8gmr}\bigr) +\nonumber
    \\  &
      \sqrt{2gmr}  K_{i\nu-1}\bigl(\sqrt{8gmr}\bigr)
    \biggr]
  \,.
  \label{eq:BsolOutside}
\end{align}
\end{subequations}
The solutions \eqref{eq:ABsolOutside} pertain to the region $r>a$, where the
potential \eqref{eq:Regularization} has the Coulomb form. For distances smaller
than the regularization distance the solution for the radial spinor components
of \eqref{eq:SpinorDef} is given in terms of cylindrical waves:
\begin{equation}
  \begin{Bmatrix}
    A(r) \\
    B(r)
  \end{Bmatrix}
  = \Ncal' \sqrt{r}
  \begin{Bmatrix}
    g J_{j-1/2}(\kt r) \\
    \kt a J_{j+1/2}(\kt r)    
  \end{Bmatrix}
  \,,
  \label{eq:ABsolInside} 
\end{equation}
where we have introduced $\kt a = \sqrt{g^2-2mga}$. The solutions 
\eqref{eq:ABsolOutside} and \eqref{eq:ABsolInside} are at energy
$\e=-m$ and need to be matched at $r=a$: 
\begin{equation}
  \Biggl( \frac{A^<(g)}{B^<(g)}\Biggr|_{r=a}
  =\Biggl(\frac{A^>(g)}{B^>(g)}\Biggr|_{r=a}
  \label{eq:Matching}
  \,.
\end{equation}
This completes the determination of the \emph{critical spinor} wavefunction,
$\Psi_c(\br)$.
Since the Dirac equation \eqref{eq:SchrodingerEq} is of first
order in the gradient operator, one is required to match only the spinor
components (and not, additionally, their derivatives as would be the case for
the Schrodinger equation), and this is enough to guarantee the continuity of the
probability current density across the region $r=a$.
The matching procedure generates a transcendental
equation for $g$, with an infinite number of solutions, on account of the
oscillating character of the functions $K_{i\nu}(z)$. For each angular momentum
$j$, the multiple solutions correspond to the values of $g$ for which the higher
bound states (i.e. those levels having the same $j$-symmetry) reach the
continuum. To determine $\gct$ we concentrate on the $s$
level ($j=1/2$), and solve Eq.~\eqref{eq:Matching}. Its smallest
solution for $g$ yields $\gct$, the renormalized critical coupling. 
%
\begin{figure}[tb]
  \centering
  \includegraphics*[clip,width=0.45\textwidth]{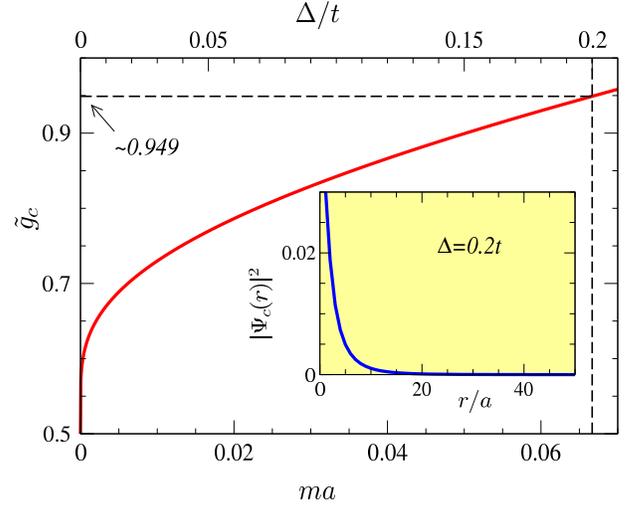}%
  \caption{(color online)
    Critical coupling, $\gct$, as a function of the mass/gap obtained from
    the solution of Eq.~\eqref{eq:Matching}.
    The top horizontal axis shows $ma$ in the units natural for the
    tight-binding calculation: $ma\to a/\lambda_C = \Delta/(3t)$.
    The inset shows the probability density associated with the critical
    wavefunction at $\e=-m$ [cfr. Eq.~\eqref{eq:Asymptotics}].
  }
  \label{fig:CriticalExact}
\end{figure}
%
A plot of $\gct$ as a function of
$ma$ is shown in Fig.~\ref{fig:CriticalExact}. 
It can be seen that the departure of $\gct$ from
the value $\gc=1/2$ is singular at the origin, which means that even an
arbitrarily small mass causes a significant change in the critical
coupling. In the figure we also point out (dashed lines) the value of
$\gct\simeq 0.949$, that corresponds to a gapped graphene spectrum with
$\Delta = 2 m\vF^2 = 0.2 t$ (in units of the tight-binding hopping parameter). 
We have chosen this value for illustration purposes, to be discussed later. The
critical probability amplitude, defined in terms of the critical wavefunction as
$\rho(r) = \Psi_c^\dagger(r)\Psi_c(r)$, is shown in the inset. The crux  of
our argument resides in the fact that $\Psi_c$ is clearly localized, as can be
inspected from the asymptotic
behavior of $\rho(r)$ for $r \gg 1/m$:
\begin{equation}
  \rho(r) = \Psi_c^\dagger(r)\Psi_c(r) 
    \sim r^{-1/2} \exp\Bigl(-2\sqrt{8 m \gct r}\Bigr)
  \,.
  \label{eq:Asymptotics}
\end{equation}
Restoring the  units, the ``Compton'' wavelength emerges as the characteristic
localization length: $m r \to m \vF r / \hbar = r/\lambda_C$.

\begin{figure}[tb]
  \centering
  \includegraphics*[clip,width=0.45\textwidth]{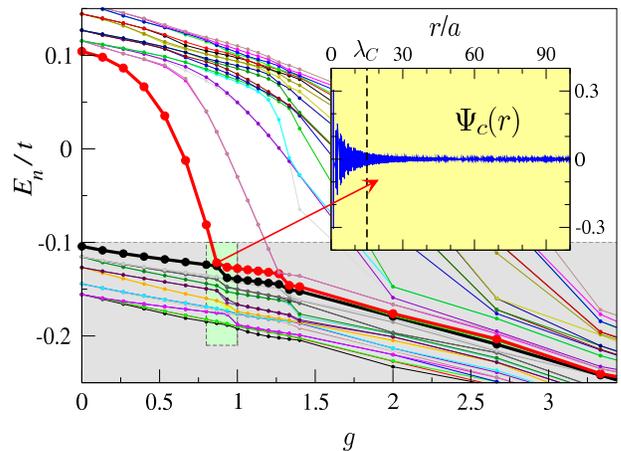}%
  \caption{(color online)
    Low energy close-up of the exact energy spectrum ($E_n$) as a 
    function of the coupling $g$ for the tight-binding problem
    \eqref{eq:H-Lattice}.
    The inset contains the exact numerical wavefunction $\Psi_c(r)$,
    with the dashed line marking $\lambda_C$ for this gap.
    The sublattice gap is $\Delta=0.2t$, and a lattice of $124^2$ sites with a
    central impurity has been used.  
  }
  \label{fig:CriticalLattice}
\end{figure}

In order to ascertain the validity of these results in the context of the
original tight-binding problem, we explicitly compare the above with the exact
numerical solution in the honeycomb lattice. 
Of particular interest are the renormalization of the critical coupling and the
character of the diving states. In Fig.~\ref{fig:CriticalLattice} we
present the exact tight-binding spectrum for a Coulomb center in the honeycomb
lattice with a sublattice energy mismatch (energy gap) of  $\Delta=0.2t$.
Inspection of the main panel reveals several features, among which: 
(i) the lowest positive level (which corresponds to $\e_0$)
immediately detaches from the continuum and sinks rapidly, followed by other
states at higher $g$;
(ii) the level $\e_0$ reaches $-m$ at $g=0.83$ and touches the lower
continuum at $g=0.87$, in a very good agreement
with the result $\gct \simeq 0.9$ predicted in
Fig.~\ref{fig:CriticalExact} from the solution of the Dirac equation
\cite{Note-Lattice};
(iii) as $g$ increases past $\gct$ this level sinks further, as is
evident from
the level avoidances highlighted by the dashed/shaded region;
(iv) the critical wavefunction, shown in the inset, is highly
concentrated around the origin, as expected from the foregoing, namely
\Eq~\eqref{eq:Asymptotics}. The lowest bound level in
Fig.~\ref{fig:CriticalLattice} is doubly degenerate.
This is expected since
although the lowest bound solution of the Dirac equation \eqref{eq:GSenergy} is
non-degenerate, there is a valley degeneracy to account for in the Dirac
description.
 
While the continuum solutions are sensitive to the finite
size of the numerical system \cite{Note-Lattice}, one does not expect the
lowest bound solution to be markedly sensitive for the sizes used in this
calculation. Hence we can extrapolate the trajectory of the lowest bound state
in Fig.~\ref{fig:CriticalLattice} to the thermodynamic limit. In that case, the
critical coupling in the lattice would be at $\simeq 0.83$, which is smaller
than the value $\gct=0.949$, and indicates that the effective cutoff distance
for the regularization \eqref{eq:Regularization} is slightly smaller than $a$.
%
     
%
%
%

%
%
\section{Vacuum Polarization and Vacuum Charging}%
\label{sec:VacuumPolarization}
It is clear, either from \Eq~\eqref{eq:Asymptotics} or from the exact
results in the lattice [Fig.~\ref{fig:CriticalLattice}(inset)], that the merging
state has a localized character. 
The characteristic length scale $\lambda_C$ is related to the gap
through $\lambda_C/a = 3 t / \Delta$. For the gap $\Delta=0.26\,  \text{eV}$,
reported in reference \cite{Zhou:2007} this corresponds to $\lambda_C \simeq 30
a$. However, the size of the merging bound state, measured as the
average radius and calculated by using the critical wave-functions, is smaller:
$\langle r \rangle = 13.9 a$.
Beyond this scale, the impurity potential is screened by essentially one charge
unit, times the product of  the spin and valley degeneracies ($N=4$). Of
course this screening is meaningfull only in a very diluted impurity
configuration ($n_i\ll 10^{12}\,\text{cm}^{-2}$ for the quoted experimental
gap).

We can appreciate this more clearly by studying the induced charge, defined as 
\begin{equation}
  \delta\rho(\br) = 
    \sum_{E\le E_F} \chi_E^\dagger(\br)\chi_E(\br) -
    \sum_{E\le -m} {\chi_E^0}^\dagger(\br){\chi_E^0(\br)}
  \,,
  \label{eq:InducedChargeDef}
\end{equation}
where $\chi_E(\br)$ are the continuum wavefunctions in the presence of 
the potential, and $\chi_E^0(\br)$ stand for the continuum wavefunctions in the
absence of potential. We work with a \emph{constant number of electrons} as the
potential is turned on, and hence $E_F \ne -m$ in general \cite{Note-Mu}. 
We need also to consider the intermediate situation in which there is a bound
state just about to merge with the lower band. The continuum wavefunctions
in this case are labeled as $\chi_E^c(\br)$. For the sake of the argument assume
that there is only one bound state that we denote by $\Psi_c(\br)$ and,
furthermore, let us accept that we can project everything onto the states $E<m$.
In other words, we accept that the states $E<m$ approximately form a complete
set in each
circumstance:
\begin{subequations}  \label{eq:Closure}
\begin{gather}
  \sum_{E\le-m} {\chi_E^0}^\dagger(\br){\chi_E^0(r')} 
    \simeq \delta(\br - \br') \,\\
  \sum_{E\le-m} {\chi_E}^\dagger(\br){\chi_E(r')} \simeq \delta(\br - \br') \,\\
  \sum_{E\le-m} {\chi_E^c}^\dagger(\br){\chi_E^c(r')} +
    {\Psi_c}^\dagger(\br)\Psi_c(r') \simeq \delta(\br - \br')
  \,.
\end{gather}
\end{subequations}
From these relations follows that, above $\gct$ when one state has dived onto
the continuum, the induced charge at
constant density defined in \eqref{eq:InducedChargeDef} can be approximated as 
\begin{equation}
  \delta\rho(\br) \simeq |\Psi_c(\br)|^2 + \delta\rho_\text{pol}(\br) -
    |\chi_{-m}^c(\br)|^2
  \,,
  \label{eq:InducedCharge}
\end{equation}
where $\delta\rho_\text{pol}(\br)$ represents the polarization charge caused by
the deformation of the continuum wavefunctions only:
\begin{equation} 
  \delta\rho_\text{pol}(\br) = \sum_{E\le -m} |\chi_E^c(\br)|^2 -
  |\chi_E^0(\br)|^2 
  \,,
  \label{eq:PolarizationCharge}
\end{equation}
with
\begin{equation} 
  \int \delta\rho_\text{pol}(\br) \, d\br = 0 \nonumber
  \,,
\end{equation}
and $\chi_{-m}^c(\br)$ represents the topmost plane wave, that appears because
we keep the number of electrons constant as $g$ is varied. Expression
\eqref{eq:InducedCharge} is valid for one bound level merging into the
continuum but can be generalized for the additional subsequent divings.

The result \eqref{eq:InducedCharge} shows that, although after the merging there
is (strictly) no localized state, the total induced charge
\eqref{eq:InducedCharge} remains essentially
determined by the profile of the critical state, $\Psi_c(\br)$, plus the
background polarization charge. The last term in eq.~\eqref{eq:InducedCharge} is
a phase-shifted wave, and thus the smallest and neglectable contribution to
$\delta\rho(\br)$.
The constant number of electrons, and the
fact that all states are normalized, evidently implies that the integral of
$\delta\rho(\br)$ over the entire volume vanishes. This allows for the
definition of $Q(R)$, the total charge inside a radius $r=R$:
\begin{equation}
  Q(R) = N \int_{|\br|<R} \delta\rho(\br) d\br
  \,.
  \label{eq:Qpol}
\end{equation}
Here $N$ represents the degeneracies of the problem: $N=4$
(spin$\times$valley) for the solution of the single-valley Dirac equation, and
$N=2$ (spin) for the numerical solution in the honeycomb lattice.
This function $Q(R)$ will raise quickly at small $R$, rapidly attaining a
maximum on account of the localized profile of $\Psi_c(\br)$ \cite{Note-Pol}. 
$Q(R)$ then returns slowly to zero as $R$ is increased further, because of the
normalization of the states and the fact that we work at a constant number of
electrons.
The maximum in $Q(R)$ (which we designate $Q_\text{max}$) is thus a
measure of
the amount of charge pulled to the vicinity of the impurity, and is our
parameter of interest. 
For $g<\gct$ the only contribution to $Q(R)$ comes from the polarization
charge, $\delta\rho_\text{pol}(\br)$; but whenever a new level dives, a
contribution of the type $|\Psi_c(\br)|^2$ adds to the total induced charge and
one should observe discontinuous jumps in $Q_\text{max}$ with increasing $g$. 

%
\begin{figure}[bt]
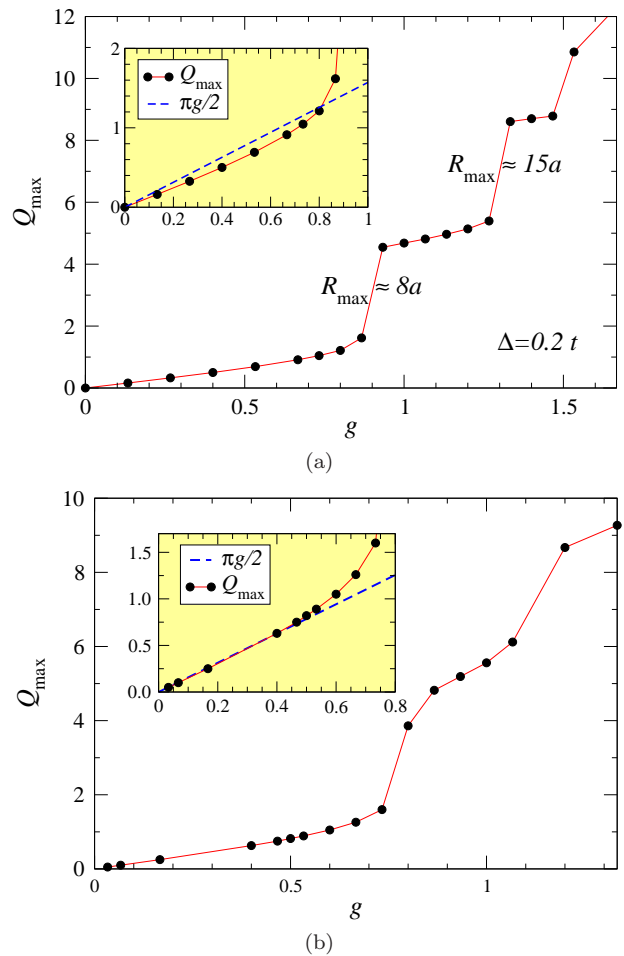

  \centering
  \subfigure[][]{%
    \includegraphics*[width=0.45\textwidth]{%
      Fig-4a.eps%
    }
    \label{fig:QpolGap}%
  }
  \subfigure[][]{%
    \includegraphics*[width=0.45\textwidth]{%
      Fig-4b.eps%
    }%
    \label{fig:Qpol}%
  }%
  \caption{(color online)
    \subref{fig:QpolGap}
    Plot of $Q_\text{max}$ (the amount of charge pulled to the vicinity 
    of the impurity, see text) versus $g$ from exact numerical
    diagonalization
    of the tight-binding problem [Fig.~\ref{fig:CriticalLattice}]. 
    The steps signal the incremental diving of bound levels
    into the lower continuum. The distances at which $Q(R)=Q_\text{max}$
    are indicated for the first two steps.
    \subref{fig:Qpol}
    The same finite system without an explicit gap ($\Delta=0$).
    In both cases the system has $124\times 124$ carbon sites and linear
    dimensions of $107a\times186a$.
    The insets amplify the low $g$ region, which is compared with the
    \ac{RPA} result for $m$=0: 
    $Q(R) = \frac{\pi}{2}g \int_{|r|<R} \delta(\br) d\br =
    \frac{\pi}{2}g$ (dashed line). 
  }
  \label{fig:VacuumCharge}
\end{figure}
%

Once again this expectation is confronted with exact results on the
lattice. We compute \eqref{eq:InducedChargeDef} numerically using the exact
wavefunctions of the tight-binding model. In Fig.~\ref{fig:QpolGap} we
plot the resulting $Q_\text{max}$. Up to $g\simeq 0.9$ the only contribution
comes from $\delta\rho_\text{pol}$ and the variation of $Q_\text{max}$ is
smooth. At the point where in Fig.~\ref{fig:CriticalLattice} the first level
merges with the continuum the first discontinuity occurs in $Q_\text{max}$,
which jumps by roughly 4 units ($N\times 1 = 4$), just as expected. Also as
expected, this charge is concentrated within a region of radius $\sim
\lambda_C$. Therefore, the approximations used to obtain expression
\eqref{eq:InducedCharge} are legitimate, and the induced charge beyond $\gct$ is
mostly determined by the profile of the critical states.

This is the precise analogue of the \ac{QED} prediction for the charging of the
vacuum. In a superheavy nucleus with high enough $Z$ the hydrogen-like bound
levels can merge with the positron continuum. When that happens, the Schwinger
mechanism of electron-positron creation becomes spontaneous since the bound
level has become degenerate with the positron states, and consequently there is
no energy cost involved in creating an electron-positron pair. The electron
occupies the bound level close to the nucleus, while the
positron escapes to infinity \cite{Greiner:1985}. Since the positron continuum 
constitutes the vacuum of QED, the spontaneous creation of an electron in
a level below $-m$ leads to a restructuring of the vacuum that acquires a charge
$Q=-2e$ \cite{Note-Charge}.
The curve in Fig.~\ref{fig:QpolGap} mimics entirely the \ac{QED}
prediction found, e.g., in Fig.~1.5 of Ref.~\onlinecite{Greiner:1985}.
The striking difference rests in the magnitude of the effect. In \ac{QED}, due
to the smallness of the real fine structure constant
($\alpha_\text{QED} \simeq 1/137$)  the nuclear charge required to reach the
critical regime is very large ($Z_c\simeq170$), and the jumps in the
polarization charge lead to an effective atomic number $Z^\text{eff}=Z_c-Q$,
with $Q\ll Z_c$: a small correction. 

In graphene, for estimate purposes, we assume a SiC substrate 
(with dielectric constant $\varepsilon \simeq 10$), as used in
Ref.~\onlinecite{Zhou:2007}, leading to $\alpha \simeq 0.4$. For the gap
$\Delta=0.26\,  \text{eV}$ seen in this experiment we obtain
$g_c=(Z\alpha)_c = 0.84$ and, consequently, the critical valence is
$Z_\text{c}\simeq 2.1$.
This means that, according to our discussion, impurities with valence
$Z\gtrsim 2$  are expected to be completely screened beyond the scale
$\langle r \rangle = 13.9a \approx 2\,\text{nm}$, since at this distance the
amount of accumulated charge in the vacuum is $Q=4$, and thus 
$Z^\text{eff}(r> \langle r \rangle) =Z- Q \approx 0$. 
In fact, for $Z<4$ over-screening takes place, and we expect 
many-body interactions to be important, effectively removing the over-screening
tendency; however such calculations are very complex and beyond
the scope of this work. In addition, we note that substrates with smaller
dielectric constants lead to higher values of $Z\alpha$
and are, in principle, capable of reducing the valences $Z$ needed to observe
the effect to $Z=1,2$. 
 
Electron-electron interactions can also lead to a change of the critical
coupling. Estimated perturbatively ($\alpha\ll 1$) this change is $\delta
g_c \sim \alpha g_c$, which would not affect significantly the physics discussed
here (but would clearly affect the precise value of $Z_c$ for a given
$\alpha$). 
We recall also that the above theory assumes the chemical potential
to be at $\mu = -m$, while the experiments of Ref.~\onlinecite{Zhou:2007} in
fact have $\mu > + m$. Tuning the Fermi level to the gap via chemical
doping or gating and varying $\alpha$ by changing the substrate's dielectric
constant would provide a possible way of exploring our predictions
experimentally.

%
%
\section{Vacuum Charging in a Finite System}%
\label{sec:FiniteSystem}
The discrete eigenstates of a finite graphene sheet, and the fact that the
density of states vanishes at the Dirac point, lead to another realization
of a gapped spectrum, and the question of whether the effects discussed above
carry to this case arises naturally. 
We have diagonalized the same lattices as above, this time without any
sublattice mismatch (i.e.: $\Delta=0$, or $m=0$), corresponding to  the massless
Dirac problem on a finite system. 
In this case we found that the gap induced by the
finite size of the system is $\Delta\simeq 0.06 t$,  $\lambda_C \approx 50 a$,
and  $\langle r \rangle \approx 21 a$.
This leads  to a renormalized $\gct \simeq 0.76$, representing a $50\%$
increase with respect to $g_c$ just from the fact that, although massless, the
system is constrained to a finite size.
For brevity we show only the integrated polarization charge $Q_\text{max}$ in
Fig.~\ref{fig:Qpol}. The curve of $Q_\text{max}$ displays the
characteristic step discontinuity at $0.7<g<0.8$, in agreement with the estimate
for $\gct$.
Of course, in this case the situation is peculiar since the effective
$\lambda_C$ is tied to the linear size of the system, $L$, and 
the gap is now $\Delta\propto L^{-1}$. Consequently  $\lambda_C$, although
still a fraction of $L$, grows with the system size.
Nevertheless, most importantly, the picture of vacuum charging discussed above
remains valid, with the vacuum charge residing at a distance $\langle r \rangle
\approx 21 a$,  significantly smaller than the linear size $L \approx 107 a$.
Thus the vacuum charging can be potentially observed in mesoscopic samples as
well.

%
%
\section{Discussion and Conclusions}%
\label{sec:Conclusions}
The fact that gapped graphene is described in terms of massive Dirac
quasiparticles makes it a rather unique solid state system. We have seen, in
particular, that effects of supercritical vacuum charging are expected when
undoped graphene is exposed to low-valence charged impurities. 

One of the key steps in our calculation is the explicit regularization of
the potential \eqref{eq:Regularization} that allows the diving of the bound
levels into the lower continuum. Since an impurity located slightly out of the
plane would still lead to a regularized Coulomb potential, our results are not
altered qualitatively because details of the regularized potential at short
distances are not important \cite{Zp:1972}. 
At the quantitative level, the regularization distance determines the
renormalization of $\gct$. Since the typical distances
to the plane expected for chemisorbed impurities in graphene are still in the
sub-nanometer range \cite{Chen:2008}, we do not expect a prohibitive increase in
$\gct$ nor, consequently, in $Z_\text{c}$.  
Therefore the charged impurity does not need to be embedded in the graphene
plane, and can be simply adsorbed to its surface. 

The theory predicts a strong tendency towards screening of the external Coulomb
potential on nanometer scales, the actual values depending on the details of the
particular situation (notably, the magnitude of the gap). We mention also that,
despite our initial assumption of attractive impurities ($g>0$), this is
only for convenience. The particle-hole symmetry of the problem ensures that the
results remain valid for repulsive charges (negative ions), in which case the
(positron) bound levels merge with the upper continuum of states at precisely
the same critical couplings.

Whereas in gapless graphene the supercritical regime is characterized by an
infinite number of resonances in the hole (positron) channel
\cite{Shytov:2007ab,Pereira:2007}, 
the phenomenon of vacuum polarization in gapped graphene is an incremental
process. Each level dives at successively higher values of the coupling
$g=Z\alpha$, leading to the the step-wise shape of the curves of $Q_\text{max}$
shown in Fig.~\ref{fig:VacuumCharge}. This translates to a quite
dissimilar screening of the supercritical impurity in the massless and
massive cases: for the latter we can have complete screening at very short
distances in a system with no carriers.

Finally, the estimates of the critical valence $Z_c\simeq 2$ made in this paper
are based on the specific parameters  of the epitaxial samples used in
Ref.~\onlinecite{Zhou:2007} (to wit the gap $\Delta$ and the dielectric
constant of the substrate). The supercritical regime is determined by 
$Z\alpha\gtrsim\gct$, to which several players contribute: $Z$ itself, the
dielectric medium (via $\alpha$), the mass/gap and the regularization distance
(via $\gct$). Some of these are more manageable to experimental control than
others. Encouraging experiments are emerging that seek control over some
of them. In Ref.~\onlinecite{Chen:2008} controlled doping with monovalent ions
has been achieved, and presumably the same could be done with divalent alkaline
ions. In Ref.~\onlinecite{Jang:2008} the fine structure constant $\alpha$
in exfoliated graphene could be controlled through changes in the dielectric
environment. And in Ref.~\onlinecite{Gruneis:2008} a gap was found for graphene
flakes on Ni surfaces. 

The charging of the vacuum in the presence of a strong Coulomb center is a long
standing prediction of \ac{QED} in strong fields that remains
unconfirmed through a direct experiment. This is an obvious consequence of the
difficulties imposed by the nuclear charge of $Z\simeq 170$ required to reach
the supercritical regime in \ac{QED}. Our calculations suggest that, under the
conditions discussed, the analogous effect might be observable in a solid-state
context, with low-valence ions.

%
%
\acknowledgments
We are  grateful to B. Uchoa, O. Sushkov, and  R. Barankov for stimulating
discussions. V.M.P. is supported by FCT via SFRH/BPD/27182/2006 and POCI 2010
via PTDC/FIS/64404/2006, and acknowledges Centro de F{\'i}sica do Porto for
computational support.


\begin{acronym}[MOSFET]
  \acro{2D}{two-dimensional}
  \acro{BZ}{Brillouin zone}
  \acro{DOS}{density of states}
  \acro{HOPG}{highly oriented pyrolytic graphite}
  \acro{LDOS}{local density of states}
  \acro{QED}{quantum electrodynamics}
  \acro{QHE}{quantum Hall effect}
  \acro{RPA}{random phase approximation}
\end{acronym}


\bibliographystyle{apsrev}
\bibliography{graphene_coulomb-2}

\begin{thebibliography}{32}
\expandafter\ifx\csname natexlab\endcsname\relax\def\natexlab#1{#1}\fi
\expandafter\ifx\csname bibnamefont\endcsname\relax
  \def\bibnamefont#1{#1}\fi
\expandafter\ifx\csname bibfnamefont\endcsname\relax
  \def\bibfnamefont#1{#1}\fi
\expandafter\ifx\csname citenamefont\endcsname\relax
  \def\citenamefont#1{#1}\fi
\expandafter\ifx\csname url\endcsname\relax
  \def\url#1{\texttt{#1}}\fi
\expandafter\ifx\csname urlprefix\endcsname\relax\def\urlprefix{URL }\fi
\providecommand{\bibinfo}[2]{#2}
\providecommand{\eprint}[2][]{\url{#2}}

\bibitem[{\citenamefont{Novoselov et~al.}(2005)\citenamefont{Novoselov, Geim,
  Morozov, Jiang, Katsnelson, Grigorieva, Dubonos, and
  Firsov}}]{Novoselov:2005}
\bibinfo{author}{\bibfnamefont{K.~S.} \bibnamefont{Novoselov}},
  \bibinfo{author}{\bibfnamefont{A.~K.} \bibnamefont{Geim}},
  \bibinfo{author}{\bibfnamefont{S.~V.} \bibnamefont{Morozov}},
  \bibinfo{author}{\bibfnamefont{D.}~\bibnamefont{Jiang}},
  \bibinfo{author}{\bibfnamefont{M.~I.} \bibnamefont{Katsnelson}},
  \bibinfo{author}{\bibfnamefont{I.~V.} \bibnamefont{Grigorieva}},
  \bibinfo{author}{\bibfnamefont{S.~V.} \bibnamefont{Dubonos}},
  \bibnamefont{and} \bibinfo{author}{\bibfnamefont{A.~A.}
  \bibnamefont{Firsov}}, \bibinfo{journal}{Nature}
  \textbf{\bibinfo{volume}{438}}, \bibinfo{pages}{197} (\bibinfo{year}{2005}).

\bibitem[{\citenamefont{{Castro Neto} et~al.}(2007)\citenamefont{{Castro Neto},
  Guinea, Peres, Novoselov, and Geim}}]{CastroNeto:2006b}
\bibinfo{author}{\bibfnamefont{A.~H.} \bibnamefont{{Castro Neto}}},
  \bibinfo{author}{\bibfnamefont{F.}~\bibnamefont{Guinea}},
  \bibinfo{author}{\bibfnamefont{N.~M.~R.} \bibnamefont{Peres}},
  \bibinfo{author}{\bibfnamefont{K.~S.} \bibnamefont{Novoselov}},
  \bibnamefont{and} \bibinfo{author}{\bibfnamefont{A.~K.} \bibnamefont{Geim}},
  \bibinfo{journal}{arXiv:0709.1163}  (\bibinfo{year}{2007}).

\bibitem[{\citenamefont{Geim and Novoselov}(2007)}]{Geim:2007}
\bibinfo{author}{\bibfnamefont{A.~K.} \bibnamefont{Geim}} \bibnamefont{and}
  \bibinfo{author}{\bibfnamefont{K.~S.} \bibnamefont{Novoselov}},
  \bibinfo{journal}{Nat. Mater.} \textbf{\bibinfo{volume}{6}},
  \bibinfo{pages}{183} (\bibinfo{year}{2007}).

\bibitem[{\citenamefont{{Castro Neto} et~al.}(2006)\citenamefont{{Castro Neto},
  Guinea, and Peres}}]{Neto:2006}
\bibinfo{author}{\bibfnamefont{A.~H.} \bibnamefont{{Castro Neto}}},
  \bibinfo{author}{\bibfnamefont{F.}~\bibnamefont{Guinea}}, \bibnamefont{and}
  \bibinfo{author}{\bibfnamefont{N.~M.~R.} \bibnamefont{Peres}},
  \bibinfo{journal}{Physics World} \textbf{\bibinfo{volume}{19}},
  \bibinfo{pages}{33} (\bibinfo{year}{2006}).

\bibitem[{\citenamefont{Katsnelson and Novoselov}(2007)}]{Katsnelson:2007}
\bibinfo{author}{\bibfnamefont{M.~I.} \bibnamefont{Katsnelson}}
  \bibnamefont{and} \bibinfo{author}{\bibfnamefont{K.~S.}
  \bibnamefont{Novoselov}}, \bibinfo{journal}{Solid State Commun.}
  \textbf{\bibinfo{volume}{143}}, \bibinfo{pages}{3} (\bibinfo{year}{2007}).

\bibitem[{\citenamefont{Zhou et~al.}(2007)\citenamefont{Zhou, Gweon, Fedorov,
  First, de~Heer, Lee, Guinea, {Castro Neto}, and Lanzara}}]{Zhou:2007}
\bibinfo{author}{\bibfnamefont{S.~Y.} \bibnamefont{Zhou}},
  \bibinfo{author}{\bibfnamefont{G.-H.} \bibnamefont{Gweon}},
  \bibinfo{author}{\bibfnamefont{A.~V.} \bibnamefont{Fedorov}},
  \bibinfo{author}{\bibfnamefont{P.~N.} \bibnamefont{First}},
  \bibinfo{author}{\bibfnamefont{W.~A.} \bibnamefont{de~Heer}},
  \bibinfo{author}{\bibfnamefont{D.-H.} \bibnamefont{Lee}},
  \bibinfo{author}{\bibfnamefont{F.}~\bibnamefont{Guinea}},
  \bibinfo{author}{\bibfnamefont{A.~H.} \bibnamefont{{Castro Neto}}},
  \bibnamefont{and} \bibinfo{author}{\bibfnamefont{A.}~\bibnamefont{Lanzara}},
  \bibinfo{journal}{Nature Materials} \textbf{\bibinfo{volume}{6}},
  \bibinfo{pages}{770} (\bibinfo{year}{2007}).

\bibitem[{\citenamefont{Nomura and MacDonald}(2006)}]{Nomura:2007}
\bibinfo{author}{\bibfnamefont{K.}~\bibnamefont{Nomura}} \bibnamefont{and}
  \bibinfo{author}{\bibfnamefont{A.}~\bibnamefont{MacDonald}},
  \bibinfo{journal}{Phys. Rev. Lett.} \textbf{\bibinfo{volume}{98}},
  \bibinfo{pages}{076602} (\bibinfo{year}{2006}).

\bibitem[{\citenamefont{Adam et~al.}(2007)\citenamefont{Adam, Hwang, Galitski,
  and Sarma}}]{Adam:2007}
\bibinfo{author}{\bibfnamefont{S.}~\bibnamefont{Adam}},
  \bibinfo{author}{\bibfnamefont{E.~H.} \bibnamefont{Hwang}},
  \bibinfo{author}{\bibfnamefont{V.~M.} \bibnamefont{Galitski}},
  \bibnamefont{and} \bibinfo{author}{\bibfnamefont{S.~D.} \bibnamefont{Sarma}},
  \bibinfo{journal}{Proc. Nat. Acad. Sci.} \textbf{\bibinfo{volume}{104}},
  \bibinfo{pages}{18392} (\bibinfo{year}{2007}).

\bibitem[{\citenamefont{Chen et~al.}(2008)\citenamefont{Chen, Jang, Fuhrer,
  Williams, and Ishigami}}]{Chen:2008}
\bibinfo{author}{\bibfnamefont{J.~H.} \bibnamefont{Chen}},
  \bibinfo{author}{\bibfnamefont{C.}~\bibnamefont{Jang}},
  \bibinfo{author}{\bibfnamefont{M.~S.} \bibnamefont{Fuhrer}},
  \bibinfo{author}{\bibfnamefont{E.~D.} \bibnamefont{Williams}},
  \bibnamefont{and} \bibinfo{author}{\bibfnamefont{M.}~\bibnamefont{Ishigami}},
  \bibinfo{journal}{Nature Physics} \textbf{\bibinfo{volume}{4}},
  \bibinfo{pages}{377} (\bibinfo{year}{2008}).

\bibitem[{\citenamefont{de~Heer et~al.}(2007)\citenamefont{de~Heer, Berger, Wu,
  First, Conrad, Li, Li, Sprinkle, Hass, Sadowski et~al.}}]{Heer:2007}
\bibinfo{author}{\bibfnamefont{W.~A.} \bibnamefont{de~Heer}},
  \bibinfo{author}{\bibfnamefont{C.}~\bibnamefont{Berger}},
  \bibinfo{author}{\bibfnamefont{X.}~\bibnamefont{Wu}},
  \bibinfo{author}{\bibfnamefont{P.~N.} \bibnamefont{First}},
  \bibinfo{author}{\bibfnamefont{E.~H.} \bibnamefont{Conrad}},
  \bibinfo{author}{\bibfnamefont{X.}~\bibnamefont{Li}},
  \bibinfo{author}{\bibfnamefont{T.}~\bibnamefont{Li}},
  \bibinfo{author}{\bibfnamefont{M.}~\bibnamefont{Sprinkle}},
  \bibinfo{author}{\bibfnamefont{J.}~\bibnamefont{Hass}},
  \bibinfo{author}{\bibfnamefont{M.~L.} \bibnamefont{Sadowski}},
  \bibnamefont{et~al.}, \bibinfo{journal}{Solid State Comm.}
  \textbf{\bibinfo{volume}{143}}, \bibinfo{pages}{92} (\bibinfo{year}{2007}).

\bibitem[{\citenamefont{Zeldovich and Popov}(1972)}]{Zp:1972}
\bibinfo{author}{\bibfnamefont{Y.~B.} \bibnamefont{Zeldovich}}
  \bibnamefont{and} \bibinfo{author}{\bibfnamefont{V.~S.} \bibnamefont{Popov}},
  \bibinfo{journal}{Sov. Phys. Usp.} \textbf{\bibinfo{volume}{14}},
  \bibinfo{pages}{673} (\bibinfo{year}{1972}).

\bibitem[{\citenamefont{Greiner et~al.}(1985)\citenamefont{Greiner, Muller, and
  Rafelski}}]{Greiner:1985}
\bibinfo{author}{\bibfnamefont{W.}~\bibnamefont{Greiner}},
  \bibinfo{author}{\bibfnamefont{B.}~\bibnamefont{Muller}}, \bibnamefont{and}
  \bibinfo{author}{\bibfnamefont{J.}~\bibnamefont{Rafelski}},
  \emph{\bibinfo{title}{Quantum Electrodynamics of Strong Fields}}
  (\bibinfo{publisher}{Springer}, \bibinfo{year}{1985}), \bibinfo{edition}{2nd}
  ed.

\bibitem[{\citenamefont{Shytov et~al.}()\citenamefont{Shytov, Katsnelson, and
  Levitov}}]{Shytov:2007ab}
\bibinfo{author}{\bibfnamefont{A.~V.} \bibnamefont{Shytov}},
  \bibinfo{author}{\bibfnamefont{M.~I.} \bibnamefont{Katsnelson}},
  \bibnamefont{and} \bibinfo{author}{\bibfnamefont{L.~S.}
  \bibnamefont{Levitov}}, \bibinfo{note}{{Phys. Rev. Lett.} \textbf{99},
  236801, 246802 (2007)}.

\bibitem[{\citenamefont{Novikov}(2007)}]{Novikov:2007}
\bibinfo{author}{\bibfnamefont{D.~S.} \bibnamefont{Novikov}},
  \bibinfo{journal}{Phys. Rev. B} \textbf{\bibinfo{volume}{76}},
  \bibinfo{pages}{245435} (\bibinfo{year}{2007}).

\bibitem[{\citenamefont{Pereira et~al.}(2007)\citenamefont{Pereira, Nilsson,
  and {Castro Neto}}}]{Pereira:2007}
\bibinfo{author}{\bibfnamefont{V.~M.} \bibnamefont{Pereira}},
  \bibinfo{author}{\bibfnamefont{J.}~\bibnamefont{Nilsson}}, \bibnamefont{and}
  \bibinfo{author}{\bibfnamefont{A.~H.} \bibnamefont{{Castro Neto}}},
  \bibinfo{journal}{Phys. Rev. Lett.} \textbf{\bibinfo{volume}{99}},
  \bibinfo{pages}{166802} (\bibinfo{year}{2007}).

\bibitem[{\citenamefont{Biswas et~al.}(2007)\citenamefont{Biswas, Sachdev, and
  Son}}]{Biswas:2007}
\bibinfo{author}{\bibfnamefont{R.~R.} \bibnamefont{Biswas}},
  \bibinfo{author}{\bibfnamefont{S.}~\bibnamefont{Sachdev}}, \bibnamefont{and}
  \bibinfo{author}{\bibfnamefont{D.~T.} \bibnamefont{Son}},
  \bibinfo{journal}{Phys. Rev. B} \textbf{\bibinfo{volume}{76}},
  \bibinfo{pages}{205122} (\bibinfo{year}{2007}).

\bibitem[{\citenamefont{Fogler et~al.}(2007)\citenamefont{Fogler, Novikov, and
  Shklovskii}}]{Fogler:2007}
\bibinfo{author}{\bibfnamefont{M.~M.} \bibnamefont{Fogler}},
  \bibinfo{author}{\bibfnamefont{D.~S.} \bibnamefont{Novikov}},
  \bibnamefont{and} \bibinfo{author}{\bibfnamefont{B.~I.}
  \bibnamefont{Shklovskii}}, \bibinfo{journal}{Phys. Rev. B}
  \textbf{\bibinfo{volume}{76}}, \bibinfo{pages}{233402}
  (\bibinfo{year}{2007}).

\bibitem[{\citenamefont{Terekhov et~al.}(2008)\citenamefont{Terekhov, Milstein,
  Kotov, and Sushkov}}]{Terekhov:2007}
\bibinfo{author}{\bibfnamefont{I.~S.} \bibnamefont{Terekhov}},
  \bibinfo{author}{\bibfnamefont{A.~I.} \bibnamefont{Milstein}},
  \bibinfo{author}{\bibfnamefont{V.~N.} \bibnamefont{Kotov}}, \bibnamefont{and}
  \bibinfo{author}{\bibfnamefont{O.~P.} \bibnamefont{Sushkov}},
  \bibinfo{journal}{Phys. Rev. Lett.} \textbf{\bibinfo{volume}{100}},
  \bibinfo{pages}{076803} (\bibinfo{year}{2008}).

\bibitem[{\citenamefont{Ando}(2006)}]{Ando:2006}
\bibinfo{author}{\bibfnamefont{T.}~\bibnamefont{Ando}}, \bibinfo{journal}{J.
  Phys. Soc. Jpn.} \textbf{\bibinfo{volume}{75}}, \bibinfo{pages}{074716}
  (\bibinfo{year}{2006}).

\bibitem[{\citenamefont{DiVincenzo and Mele}(1984)}]{DiVincenzo:1984}
\bibinfo{author}{\bibfnamefont{D.~P.} \bibnamefont{DiVincenzo}}
  \bibnamefont{and} \bibinfo{author}{\bibfnamefont{E.~J.} \bibnamefont{Mele}},
  \bibinfo{journal}{Phys. Rev. B} \textbf{\bibinfo{volume}{29}},
  \bibinfo{pages}{1685} (\bibinfo{year}{1984}).

\bibitem[{\citenamefont{Abramowitz and Stegun}(1964)}]{Abramowitz:1964}
\bibinfo{author}{\bibfnamefont{M.}~\bibnamefont{Abramowitz}} \bibnamefont{and}
  \bibinfo{author}{\bibfnamefont{I.~A.} \bibnamefont{Stegun}},
  \emph{\bibinfo{title}{Handbook of Mathematical Functions}}
  (\bibinfo{publisher}{Dover}, \bibinfo{address}{New York},
  \bibinfo{year}{1964}).

\bibitem[{\citenamefont{Landau and Lifshitz}(1981)}]{Landau-QM:1981}
\bibinfo{author}{\bibfnamefont{L.~D.} \bibnamefont{Landau}} \bibnamefont{and}
  \bibinfo{author}{\bibfnamefont{E.~M.} \bibnamefont{Lifshitz}},
  \emph{\bibinfo{title}{Quantum Mechanics: Non-Relativistic Theory}}
  (\bibinfo{publisher}{Pergamon Press}, \bibinfo{year}{1981}).

\bibitem[{\citenamefont{Case}(1960)}]{Case:1960}
\bibinfo{author}{\bibfnamefont{K.~M.} \bibnamefont{Case}},
  \bibinfo{journal}{Phys. Rev.} \textbf{\bibinfo{volume}{80}},
  \bibinfo{pages}{797} (\bibinfo{year}{1960}).

\bibitem[{\citenamefont{Khalilov and Ho}(1998)}]{Khalilov:1998}
\bibinfo{author}{\bibfnamefont{V.~R.} \bibnamefont{Khalilov}} \bibnamefont{and}
  \bibinfo{author}{\bibfnamefont{C.-L.} \bibnamefont{Ho}},
  \bibinfo{journal}{Mod. Phys. Lett. A} \textbf{\bibinfo{volume}{13}},
  \bibinfo{pages}{615} (\bibinfo{year}{1998}).

\bibitem[{\citenamefont{Popov}(1971)}]{Popov:1971}
\bibinfo{author}{\bibfnamefont{V.~S.} \bibnamefont{Popov}},
  \bibinfo{journal}{Sov. J. Nucl. Phys.} \textbf{\bibinfo{volume}{12}},
  \bibinfo{pages}{235} (\bibinfo{year}{1971}).

\bibitem[{\citenamefont{M\"uller and Rafelski}(1975)}]{Muller:1975}
\bibinfo{author}{\bibfnamefont{B.}~\bibnamefont{M\"uller}} \bibnamefont{and}
  \bibinfo{author}{\bibfnamefont{J.}~\bibnamefont{Rafelski}},
  \bibinfo{journal}{Phys. Rev. Lett.} \textbf{\bibinfo{volume}{34}},
  \bibinfo{pages}{349} (\bibinfo{year}{1975}).

\bibitem[{Not({\natexlab{a}})}]{Note-Lattice}
\bibinfo{note}{In Fig.~\ref{fig:CriticalLattice} the boundaries of the
  continuum also change with $g$. This is due to the fact that, in such a
  finite system as the one analyzed there, all the energy levels are
  phase-shifted by the potential. This remains true for the levels close to the
  gap and therefore the merging of the bound solution with the continuum in a
  finite-sized lattice appears at a larger coupling. This phase shift effect
  vanishes when the size of the system becomes infinite.}

\bibitem[{Not({\natexlab{b}})}]{Note-Mu}
\bibinfo{note}{The chemical potential (the last occupied level) follows the
  bold black line in Fig.~\ref{fig:CriticalLattice}}.

\bibitem[{Not({\natexlab{c}})}]{Note-Pol}
\bibinfo{note}{There is an additional localized contribution that comes from
  $\delta\rho_\text{pol}$ which is reminiscent of the localized screening
  charge $\propto\delta(\br)$ in the massless case \cite{Terekhov:2007}. This
  term yields the smoothly varying part in Fig.~\ref{fig:VacuumCharge}, but is
  not our focus here.}

\bibitem[{Not({\natexlab{d}})}]{Note-Charge}
\bibinfo{note}{The factor of 2 comes from the spin degeneracy.}

\bibitem[{\citenamefont{Jang et~al.}(2008)\citenamefont{Jang, Adam, Chen,
  Williams, Sarma, and Fuhrer}}]{Jang:2008}
\bibinfo{author}{\bibfnamefont{C.}~\bibnamefont{Jang}},
  \bibinfo{author}{\bibfnamefont{S.}~\bibnamefont{Adam}},
  \bibinfo{author}{\bibfnamefont{J.-H.} \bibnamefont{Chen}},
  \bibinfo{author}{\bibfnamefont{E.~D.} \bibnamefont{Williams}},
  \bibinfo{author}{\bibfnamefont{S.~D.} \bibnamefont{Sarma}}, \bibnamefont{and}
  \bibinfo{author}{\bibfnamefont{M.~S.} \bibnamefont{Fuhrer}},
  \bibinfo{journal}{arXiv:0805.3780}  (\bibinfo{year}{2008}).

\bibitem[{\citenamefont{Gruneis and Vyalikh}(2008)}]{Gruneis:2008}
\bibinfo{author}{\bibfnamefont{A.}~\bibnamefont{Gruneis}} \bibnamefont{and}
  \bibinfo{author}{\bibfnamefont{D.~V.} \bibnamefont{Vyalikh}},
  \bibinfo{journal}{Phys. Rev. B} \textbf{\bibinfo{volume}{77}},
  \bibinfo{pages}{193401} (\bibinfo{year}{2008}).

\end{thebibliography}

\end{document}